\documentclass{kapproc} 

\usepackage{procps} 

\usepackage[dvips]{graphicx}

 \chapsectrunningheads

\upperandlowercase

\kluwerbib 

\begin{document}



\articletitle[The Molecular ISM of Low Surface\\ Brightness 
Spiral Galaxies]{The Molecular ISM of Low Surface\\ Brightness Spiral Galaxies}

\chaptitlerunninghead{Molecular ISM of LSB Galaxies}




 \author{L. D. Matthews}
 \affil{Harvard-Smithsonian Center for Astrophysics, Cambridge, MA USA}
 \email{lmatthew@cfa.harvard.edu}















\begin{abstract}
I summarize some results from the recent CO survey of late-type, low surface
brightness (LSB) spiral galaxies by Matthews et al. (2005). We have
now detected CO emission from six late-type, LSB
spirals, demonstrating that despite their typical low metallicities
and low mean gas surface densities, 
some LSB galaxies contain a molecular
medium that is traced by CO. We find that the
CO-detected LSB spirals adhere to the same $M_{\rm H_{2}}$-FIR correlation as
brighter  galaxies. We also find a significant
drop-off in the detectability of CO among
low-to-intermediate surface brightness galaxies with 
$V_{\rm rot}{~\rlap{$<$}{\lower 1.0ex\hbox{$\sim$}}}$90km s$^{-1}$,
pointing toward fundamental
changes in the physical conditions of the ISM with decreasing disk mass.
\end{abstract}


\section{Background}
Low surface brightness (LSB) 
spiral galaxies are defined as having central disk surface brightnesses
$\mu(0)_{B}${~\rlap{$>$}{\lower 1.0ex\hbox{$\sim$}}}23
mag arcsec$^{-2}$, indicating low stellar surface densities. 
However, despite their faint optical appearances, the majority of 
LSB spirals show
evidence for ongoing star formation, including blue colors, 
H$\alpha$ emission and/or resolved populations of young stars
(e.g., Schombert et al. 1992; Gallagher \& Matthews
2002).  
Signatures of star formation are frequently coupled with
large atomic gas fractions ($M_{\rm HI}/L_{B}\sim$1), underscoring that LSB spirals
are not the faded remnants of brighter galaxies. 
In a number of cases, evidence is also seen for stellar populations
spanning a wide range of ages (e.g., van den
Hoek et al. 2000), implying that LSB spirals 
have been forming stars for a significant fraction of a Hubble
time---but with low efficiency. 

Lingering questions regarding the evolutionary history of LSB spirals
are {\it why} their star formation 
has remained suppressed, and what physical processes regulate the type of
low-level star formation that is observed in these systems.  Given that
LSB spirals comprise a
significant fraction ($\sim$50\%) of the local disk galaxy population
(e.g., Minchin et al. 2004), answers to these questions are crucial for our overall
understanding of the star formation history of the universe.  In
addition, they are relevant to understanding star formation
in other low-density and/or low-metallicity environments, including
protogalaxies, the outskirts of giant galaxies, and damped Ly$\alpha$
absorption systems.

\section{Studies of the ISM of LSB Spirals}
\subsection{Past Results}
Key to understanding the process of star formation in LSB spirals is an
improved knowledge of the composition and structure of their
interstellar medium (ISM). To date,
most of our knowledge of the ISM in LSB spirals comes from 
studies of their {\mbox{H\,{\sc i}}} gas, which appears to be by far the dominant
component of their ISM ({~\rlap{$>$}{\lower 1.0ex\hbox{$\sim$}}}95\%
by mass). {\mbox{H\,{\sc i}}} studies have
shown that while {\mbox{H\,{\sc i}}} is typically present throughout the stellar disk
of LSB spirals, {\mbox{H\,{\sc i}}} surface densities
often fall below the critical threshold for instability-driven star formation
throughout most or all of their disks (e.g., van der
Hulst et al. 1993; de Blok et al. 1996). 

In a broad sense, the low {\mbox{H\,{\sc i}}} densities of LSB spirals
seem to 
account naturally 
for their inefficient star formation. However, this explanation is not entirely
satisfactory for several reasons. First, important ISM parameters
including the gas scale height,  volume density, and turbulent
velocity remain poorly constrained. Secondly, we know some
star formation is occurring in LSB spirals in spite of subcritical 
{\mbox{H\,{\sc i}}}
surface densities. Furthermore, it is ultimately
from the molecular, not the atomic gas that stars form. It is clear that
a more comprehensive picture of star formation in LSB spirals
requires a more complete knowledge of their multi-phase ISM.

Unfortunately, direct searches for molecular gas in LSB spirals
have proved to be challenging. Indeed,
initial searches for CO emission from
late-type, LSB spirals failed to yield any detections (Schombert et
al. 1990; de Blok \& van der Hulst 1998), leading to the
suggestions that the
metallicities of typical LSB spiral may be too low for the formation
of CO molecules or efficient cooling of the gas, 
that the interstellar pressures may be insufficient to
support molecular clouds, or  that star formation may occur
directly from the atomic medium (Schombert et al. 1990; Bothun et
al. 1997; Mihos et al. 1999). 

\subsection{New Results from Deep CO Surveys}
To partially circumvent the challenges of detecting CO emission from
LSB spirals, my collaborators and I began surveying examples of LSB
spirals viewed {\it edge-on}, using observations 2-3 times deeper than
previous studies. Our targets comprised extreme late-type (Scd-Sm)
LSB spirals with redshifts
$V_{r}\le$2000 km s$^{-1}$. Two advantages of our approach are: (1) the
column depth of the molecular gas is enhanced for an edge-on geometry;
and (2) an edge-on viewing angle allows complementary studies of the vertical
structure of various ISM and stellar components of the galaxies at other wavelengths. 

\begin{figure}[h]
\includegraphics[height=1.75in]{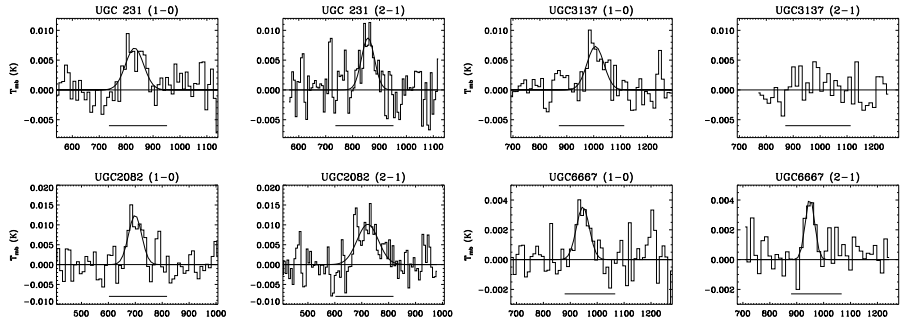}
\vspace{-1.0cm}
\caption{CO(1-0) and (2-1) spectra of four late-type, LSB spiral
galaxies detected by M05.\protect\label{fig:spectra} }
\end{figure}

In a pilot
survey with the NRAO 12-m telescope, we
detected for the first time $^{12}$CO(1-0) emission from
three late-type, LSB spiral galaxies (Matthews \& Gao 2001). More
recently, we followed up
with a more extensive survey of 15 LSB spirals in both the $^{12}$CO(1-0) and
$^{12}$CO(2-1) lines using the IRAM 30-m telescope (Matthews et
al. 2005; hereafter M05). In this latter survey, we
detected CO emission from the nuclear regions of four LSB
spirals, one of which was previously detected by Matthews \& Gao
(2001; Fig.~\ref{fig:spectra}). For the galaxies detected in these two
surveys, we estimate the molecular hydrogen content of
the nuclear regions (central 1-3~kpc) to be
$M_{\rm H_{2}}\approx(0.3-2)\times10^{7}~M_{\odot}$, assuming a
standard Galactic
CO-to-H$_{2}$ conversion factor. While the conversion of CO flux to
H$_{2}$ mass in low-density, low-metallicity galaxies
can be rather uncertain,
{\it these observations have clearly established
that at least some bulgeless, late-type, LSB spirals contain modest
amounts of molecular gas in their nuclear regions, 
and  that CO traces at least some
fraction of this gas.} In addition, our results establish that a bulge
is not a prerequisite for the presence of molecular gas at the centers
of low-density LSB galaxies. Therefore our surveys extend the realm of
CO-detected LSB spirals from the giant, bulge-dominated LSB systems
detected by O'Neil et al. (2000,2003) and O'Neil \& Schinnerer (2004) 
to the more common, low-mass, pure-disk LSB systems.

While the samples of late-type, LSB spirals surveyed in CO are still
small, already some interesting trends are emerging.
Here I briefly describe two of our key findings. For further results and
discussion, I refer the reader to M05.

\subsubsection{LSB Spirals and the FIR-H$_{2}$ Correlation}
For bright, massive spiral galaxies, there is a well-established
correlation between far-infrared (FIR) luminosity and H$_{2}$ mass (or
CO luminosity; e.g., Young \&
Scoville 1991). This correlation
is assumed to arise from heating of dust grains embedded in
giant molecular clouds (GMCs) by hot young stars. 
There are a number of reasons why this correlation
might break down for LSB spirals; for example, 
if molecular gas in LSB spirals resides primarily outside 
GMCs, if their stellar mass function is biased toward low-mass stars
(Lee et al. 2004), or if the appropriate CO-to-H$_{2}$
conversion factor for these galaxies is very different from what we
have assumed. 

Fig.~\ref{fig:FIR} shows a plot of the nuclear H$_{2}$ masses (or
3$\sigma$ upper limits) for
our CO survey galaxies, versus the FIR luminosity derived from {\it
IRAS} data. Also plotted  is a sample
of extreme late-type spirals recently surveyed in CO by Boeker et
al. (2003; hereafter B03).
The B03 sample comprises the same range of redshifts and Hubble
types as our LSB spiral CO surveys, but
covers  a wide range in surface brightness, including two
LSB spirals [$\mu_{I}(0)\ge21.4$~mag arcsec$^{-2}$], 15 intermediate
surface brightness (ISB) spirals [$18.7\le\mu_{I}(0)<21.4$~mag arcsec$^{-2}$], and 25
high surface brightness (HSB) spirals. Also overplotted as a solid
line on Fig.~\ref{fig:FIR} is
the H$_{2}$-FIR relation derived by B03 for a sample
of brighter, more massive galaxies. 

Fig.~\ref{fig:FIR} reveals that
the CO-detected LSB spirals delineate a
remarkably tight extension of the H$_{2}$-FIR defined by brighter
galaxies. Only a handful of the LSB/ISB 
galaxies undetected in CO show  evidence of possible deviation from this correlation.
Our findings
suggest that as in brighter galaxies, the CO detected in LSB spirals is
associated with dense molecular clouds and sights of  massive
star formation rather than a more diffuse molecular medium. 

\subsubsection{A Link between Disk Rotational Velocity and the Detectability of CO in
Low-Mass Galaxies}
\begin{figure}[h]
\includegraphics[height=3.75in,angle=90]{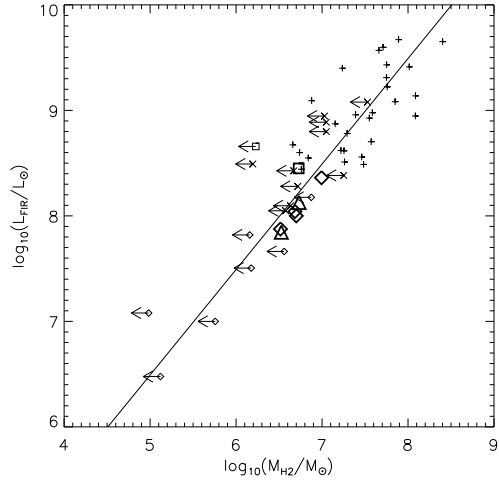}
\vspace{-0.3cm}
\caption{Log of the far-infrared luminosity versus log of the 
nuclear H$_{2}$
mass for extreme late-type spiral galaxies. $M_{\rm H_{2}}$ was
derived from a single measurement with a telescope
beam subtending $\sim$0.2-2~kpc toward each galaxy's center.
Open symbols are LSB spirals taken from three different sources:
triangles (Matthews \& Gao 2001); diamonds (M05);
squares (B03). `+' symbols are HSB spirals and `X'
symbols are ISB spirals, both taken from B03.\protect\label{fig:FIR} }
\end{figure}

\begin{figure}[h]
\includegraphics[height=3.75in,angle=90]{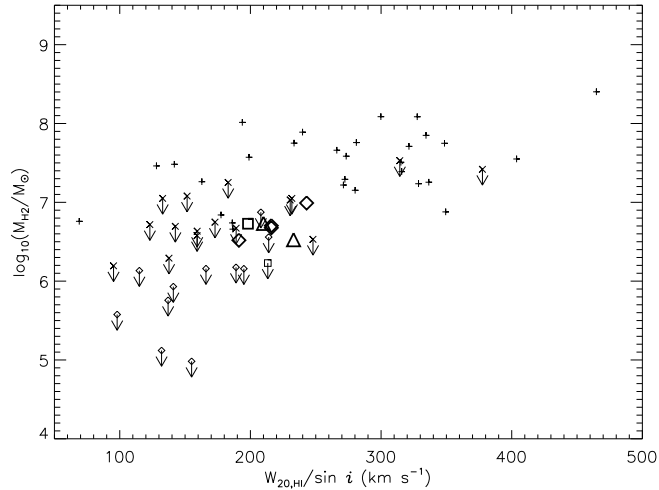}
\vspace{-0.3cm}
\caption{Log of the nuclear 
molecular hydrogen mass versus inclination-corrected
total {\mbox{H\,{\sc i}}} linewidth for extreme late-type spirals. 
Symbols are as in Figure~\ref{fig:FIR}.\protect\label{fig:W20}}
\end{figure}

Fig.~\ref{fig:W20} plots the nuclear H$_{2}$ mass versus the
inclination-corrected total
{\mbox{H\,{\sc i}}} linewidth ($W_{20}/{\rm sin}i$) 
for the same samples shown in Fig.~\ref{fig:FIR}. 
$W_{20}$ is related to
the maximum rotational velocity of disk galaxies as
$V_{\rm rot}\approx0.5(W_{20} - 20)/(2{\rm sin}i)$ (see M05). 

We see that for larger rotational velocities ($W_{20}/{\rm sin}i >250$
km s$^{-1}$),  the quantities plotted on Fig.~\ref{fig:W20} shows a nearly flat
correlation. However, for galaxies with $W_{20}/{\rm sin}i <200$ km s$^{-1}$
(corresponding to 
$V_{\rm rot}{~\rlap{$<$}{\lower 1.0ex\hbox{$\sim$}}}$90 km s$^{-1}$), the 
inferred nuclear H$_{2}$ mass (i.e., the CO detectability) begins
to drop
significantly, and no ISB or LSB spirals 
below this limit have so far been detected in CO. As discussed by M05, 
it appears that neither decreasing mean {\mbox{H\,{\sc
i}}} surface
density, nor decreasing metallicity among
the lower mass galaxies can fully account for this trend. 
Fig.~\ref{fig:W20} therefore points toward a decreasing
concentration of molecular gas in the inner regions of galaxies with decreasing
rotational velocity.  This trend seems to depend only weakly on
optical central surface brightness in the sense that both LSB and ISB
galaxies show similar declines in CO detectability with $V_{\rm rot}$,
although a few HSB systems with low $V_{\rm rot}$ do have CO detections.

It is interesting to note that the
characteristic velocity, $V_{\rm rot}\approx$90 km s$^{-1}$, below
which we
see a decline in the detectability of CO among  late-type
spirals is similar to the 
velocity characterizing a slope change in the 
optical Tully-Fisher relation for {\mbox{H\,{\sc i}}}-rich, extreme
late-type disks found by Matthews et al. (1998) ($V_{\rm rot}\sim$90
km s$^{-1}$), the characteristic rotational
velocity below which star formation appears to have been
suppressed at high redshift ($V_{\rm rot}\sim$100 km s$^{-1}$; Jimenez
et al. 2005), and the rotational velocity
below which galaxy dust lanes are seen to disappear ($V_{\rm
rot}\sim$120 km s$^{-1}$; Dalcanton et al. 2004).  {\it All of these results are
consistent with fundamental changes in some key parameter(s) governing the ISM
conditions and the regulation of star formation in
low-mass galaxies}. Possible causes could include: a
decreasing fraction the disk unstable to GMC formation (e.g., Li et
al. 2005), a change in the characteristic velocity for turbulence
(Dalcanton et al. 2004),  an increasing fraction of the disk below the
critical pressure needed for the existence of a cold ISM phase
(Elmegreen \& Parravano 1994),  the
increasing dominance of gas relative to stars in the underlying disk
potential,  and/or decreasing effects of rotational shear
(e.g., Martin \& Kennicutt 2001). We are presently undertaking
additional multiwavelength observations of our CO survey sample to better
constrain the importance of these various effects.

\vspace{-0.5cm}

\vspace{-0.3cm}
 


%




\end{document}